\documentclass[a4,aps,twocolumn,showpacs,amsmath,floatfix]{revtex4}
\usepackage{graphicx}
\usepackage{dcolumn}
\usepackage{color}
\usepackage{amsmath}
\usepackage{here}
\begin{document}
\title{Stable bright solitons in two-component Bose-Einstein condensates}

\author{A.I. Yakimenko$^{1,2}$, K.O. Shchebetovska$^1$, S.I. Vilchinskii$^1$, M. Weyrauch$^3$}

\affiliation{$^1$ Department of Physics, Taras Shevchenko National University,
Kiev 03022, Ukraine\\
$^2$
Institute for Nuclear Research, Kiev 03680, Ukraine\\
$^3$ Physikalisch-Technische Bundesanstalt, Bundesallee 100, D-38116 Braunschweig, Germany}

\begin{abstract}
Two-dimensional (2D) fundamental soliton-soliton pairs are
investigated in binary mixtures of Bose-Einstein condensates with
attractive interactions between atoms of the same type. Both
attractive and repulsive interactions between atoms of different
type are considered. The general properties of the stationary states
are investigated variationally and numerically, and the stability
regions of the soliton-soliton pairs are determined.
\end{abstract}

 \maketitle

\section{Introduction}

Binary mixtures of Bose-Einstein condensates (BECs) are interacting
quantum systems of macroscopic scale which exhibit rich physics not
accessible in a single-component degenerate quantum gas. They open
up intriguing possibilities for a number of important physical
applications, including quantum simulation~\cite{qubit}, quantum
interferometry \cite{Zawadzki2010}, and precision
measurements~\cite{Sci10,BhongalePRL08}. Experimentally,
multi-component BECs are generated as mixtures of atoms in different
hyperfine states or by simultaneously trapping different atomic
species. The key difference between multi-component and
single-component BECs is the inter-component interaction. It is significant that the
strength and even the sign of the various atomic interactions can be
controlled by means of the Feshbach resonance~\cite{exp1,exp2}.

Various types of nonlinear matter wave structures have been
predicted for multi-component
BECs~\cite{darkgrey,darkdark,wall2,wall3,symbiotic}. However, many
basic properties of two-component BECs remain to be investigated
even in simple mean-field approaches based on coupled Gross-Pitaevskii equations (GPEs) with a vector order parameter. Most
previous work on two-dimensional (2D) and three-dimensional (3D)
vector solitonic structures focused on BECs with repulsive
intra-component
interactions~\cite{MalomedJPhB00,Skryabin,Garcia1,Garcia2,AdhikariPLA05}.

As is well known, bright 2D and 3D solitons, described by a GPE with
attractive nonlinearity, are unstable and may collapse in a finite
amount of time, if the number of atoms in the condensate exceeds a
critical value (see e.g. Ref.~\cite{BergeKivshar2000}). Recent
theoretical investigations \cite{PRA2009} predicted the existence of
stable soliton-vortex pairs in trapped BECs with attractive
intra-component interactions. However, the ground state 2D soliton-soliton structures
in such BECs have never been
investigated and will be the focus of the present work. Note, that
in a different context soliton-soliton pairs, described by a similar model, but without linear potential, are found to be unstable
\cite{PRE2010}. It is reasonable to expect that a stabilization of
fundamental vector solitons could occur in an additional external
trapping potential.

In the present work we perform a detailed theoretical analysis of
nonlinear matter-wave structures in binary mixtures of atoms with
attractive intra-component interactions and attractive as well as
repulsive inter-component interactions. General properties of the
steady states of such systems are investigated by means of a
variational analysis and numerical simulations. The conditions for
the existence and stability of matter-wave vector solitons are
revealed.

\section{Model}
We consider a binary mixture of BECs at zero temperature described
in mean-field approximation by two coupled Gross-Pitaevskii
equations (GPEs) ($j \in \{1,2\}$)
 \begin{equation}
\label{GP1}
 i\hbar\frac{\partial \Psi_j}{\partial t}=\left[\hat{H}_j+g_{jj}|\Psi_j|^2+g_{j, 3-j}|\Psi_{3-j}|^2\right]\Psi_j,
\end{equation}
with $\hat{H}_j=-\frac{\hbar^2}{2M_j}\nabla^2+V_j(\mathbf{r})$, and
$M_j$ the mass of an atom of type $j$ loaded into the
axially-symmetric harmonic external trapping potential
$V_j(\mathbf{r})=M_j\omega_{\perp}^{2}(x^{2}+y^{2})/2+M_j\omega_z^2z^2/2$.
Interactions between atoms of the same type ({\it intra}-component
interactions) are characterized by the diagonal coupling
coefficients $g_{jj}=4\pi\hbar^2 a_{jj}/M_j$, where $a_{jj}$ are the
$s$-wave scattering lengths for binary collisions between these
atoms. {\it Inter}-component interactions are controlled by the
off-diagonal coupling terms  $g_{12}=g_{21}=2\pi\hbar^2 a_{12}/M_*$
with the reduced mass $M_*=M_1M_2/(M_1+M_2)$. Note, that our simple
mean field approach provides a reasonable approximation for the
evolutionary scenarios investigated here, however, a mean-field
approach of course fails at the final stage of the collapse when the
atomic density increases catastrophically.

We assume that the longitudinal trapping frequency $\omega_z$ is
much larger than the transversal trapping frequency $\omega_\perp$
($\omega_z\gg \omega_\perp$), and that the nonlinear interactions
$\sim g_{ij}$ are weak with respect to the confinement strength of
the potential in the longitudinal direction.  In this case the BEC
is ``disk-shaped" and we may assume that the longitudinal motion of
condensates is frozen in,
$\Psi_j(\mathbf{r},t)=\tilde\Psi_j(x,y,t)\Upsilon_j(z,t),$ where
$\Upsilon_j(z,t)=(l_{zj}\sqrt{\pi})^{-1/2}\exp(-\frac{i}{2}\omega_zt-\frac12z^2/l_{zj}^2)$
is the ground state wave function of the longitudinal harmonic
trapping potential, $l_{zj}=\sqrt{\hbar/(M_j\omega_{z})}$. After
integrating out the longitudinal coordinates, the GP equations take
the effective 2D form
\begin{equation}\label{GP2D}
i\frac{\partial \tilde\Psi_j}{\partial t}+\left(\hat h_j+|\tilde\Psi_j|^2
+b_{j,3-j}|\tilde\Psi_{3-j}|^2\right)\tilde\Psi_{j}=0,
\end{equation}
where $\hat h_1=\Delta_\perp-r^2$, $\hat
h_2=\kappa\Delta_\perp-\frac{r^2}{\kappa}$, $\kappa=M_1/M_2$,
$r=\sqrt{x^2+y^2}$, and $\Delta_\perp=\partial^{2}/\partial
x^{2}+\partial^{2}/\partial y^{2}$ is the 2D Laplacian. Here we have
introduced the dimensionless variables $(x,y)\to (x,y)/l_{\perp1}$,
$t\to t/\tau$, and $\tilde\Psi_j\to \tilde\Psi_j/\sqrt{C_j}$,  where
$l_{\perp1}=\sqrt{\hbar/(M_1\omega_\perp)}$, $\tau=2/\omega_\perp$,
$C_j=\hbar\omega_\perp\sqrt{\pi(l_{z1}^2+l_{z2}^2)}/(2|g_{jj}|)$.
Dimensionless coupling parameters are defined by
$b_{12}=-\frac{g_{12}}{|g_{11}|}\sqrt{\frac{2}{1+\kappa}}$ and
$b_{21}=-\frac{g_{21}}{|g_{22}|}\sqrt{\frac{2\kappa}{1+\kappa}}$.
For simplicity, we consider here only the symmetric case $M_1=M_2$
and $g_{11}=g_{22}$, so that $\kappa=1$ and $b_{12}=b_{21}=\sigma$.
Furthermore, we specifically consider only cases where the diagonal
part of the interaction matrix $g_{ij}$ is attractive ($g_{11}<0$,
$g_{22}<0$) while the off-diagonal terms $g_{12}=g_{21}$ may be
attractive ($g_{12}<0, \sigma>0 $) or repulsive ($g_{12}>0,
\sigma<0$).

The coupled differential equations (\ref{GP2D}) have the following integrals of motion: \\
(i) the number of particles in each component
\begin{equation}
N_j=\int |\tilde\Psi_j|^2d^2\textbf{r},
\end{equation}
(ii) the total energy
\begin{equation}\label{E}
E=E_1+E_2-\sigma\int{|\tilde\Psi_1|^2|\tilde\Psi_2|^2d^2\textbf{r}},
\end{equation}
with
$$
E_j=\int\left\{|\nabla\tilde\Psi_j|^2+r^2|\tilde\Psi_j|^2-
\frac{1}{2}|\tilde\Psi_j|^4\right\} d^2\textbf{r},
$$
(iii) momentum, and (iv) angular momentum.

\section{Stationary solutions}
We consider now stationary vector soliton solutions of the coupled GPEs (\ref{GP2D}). We are looking
for a ground state of the form
%
\begin{equation}
\label{Psi} \tilde\Psi_j(\mathbf{r},t)=\psi_j(r)e^{-i\mu_j t}
\end{equation}
where $r=\sqrt{x^2+y^2}$. Each solution is characterized by chemical potentials $\mu_j$. The real functions $\psi_j(r)$ satisfy
the coupled equations
\begin{equation}\label{StationaryEqs}
\mu_j \psi_j +
\psi''_j+\frac{1}{r}\psi'_j-r^2\psi_j+\left[\psi_j^2+\sigma
\psi_{3-j}^2\right]\psi_j=0
\end{equation}
and the boundary conditions $\psi_j'(0)=0$ and $\psi_j(\infty)=0$.
At fixed strength of the inter-component interaction $\sigma$ we
then obtain a two-parameter family of vector soliton solutions (with
the chemical potentials $\mu_1$ and $\mu_2$ as parameters).

Obviously, vector solitons with the same chemical potential
$\mu_1=\mu_2=\tilde\mu$ and the same radial profiles $\psi_j(r)$ are
described by a single GPE with a harmonic potential and an effective
interaction strength  $\tilde\sigma=1+\sigma$. It is known (see e.g.
Ref.~\cite{BergePRE02}) that 2D solitonic solutions of this equation
exist for $\tilde\mu<2$ if $\tilde\sigma>0$ and for $\tilde\mu>2$ if
$\tilde\sigma<0$. The value $\tilde\mu=2$ coincides with the
eigenvalue of the ground state of a linear Schr\"{o}dinger equation with
harmonic oscillator potential, i.e.
nonlinear terms vanish and the number of atoms tends to zero  when
$\tilde\mu\to 2$. If $\tilde\sigma<0$, the number $N$ of atoms grows
rapidly for $\tilde\mu>2$; if $\tilde\sigma>0$, $N$ saturates at a
critical value $N_{\rm cr}/\tilde\sigma$ where $\tilde\mu\to
-\infty$. Here, $N_{\rm cr}=11.68$ is the number of atoms in the
fundamental Townes soliton, which is the localized solution of a
single GPE without external trap.

Using these results for one-component solitons as a guide, one may
expect qualitatively different properties for vector solitons with
strong repulsion ($\sigma<-1$) or weak repulsion $-1<\sigma<0$.
Analogously, we will also discuss separately the properties of the
soliton-soliton pairs with weak attraction ($0<\sigma<1$) and strong
attraction ($\sigma>1$).

\begin{figure}
\includegraphics[width=3.4in]{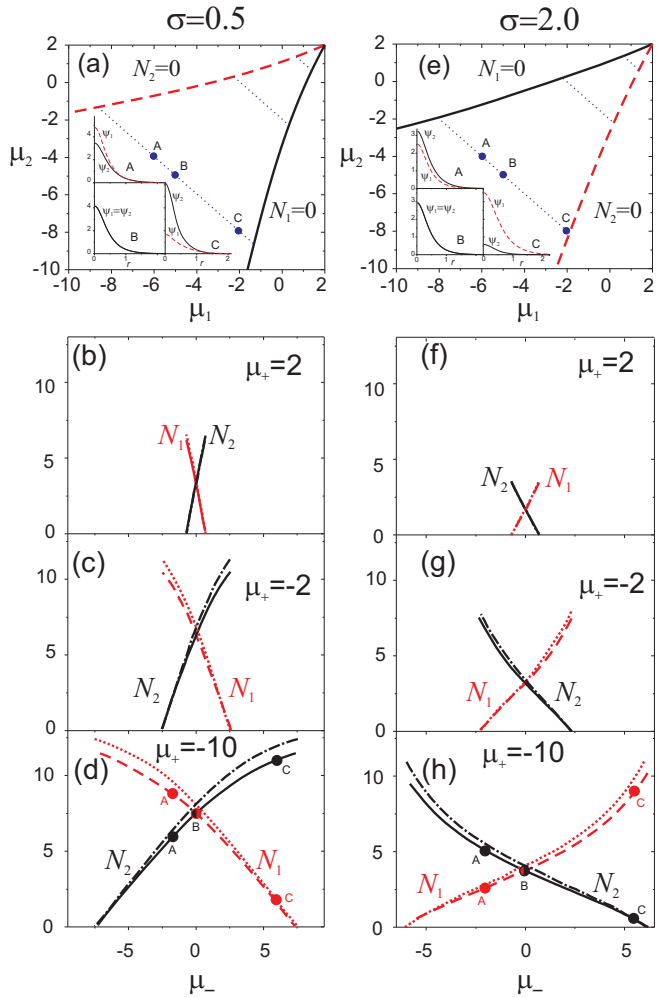}
\caption{(Color online)  (a),~(e) Existence region for vector
solitons in the $(\mu_1,~\mu_2)$ plane for $\sigma=0.5$ and
$\sigma=2.0$. The insets show examples for radial profiles
$\psi_1(r)$ (red dashed line) and $\psi_2(r)$ (black solid line) at points $A$, $B$, and $C$
in the $(\mu_1,~\mu_2)$ plane: $A= (-6, -4)$, $B= (-5,-5)$, $C=(-2, -8)$.
(b)-(d), (f)-(h) The number of atoms $N_1$ (red dashed line for numerical,
red dotted line for variational results) and $N_2$ (black solid line
for numerical results, black dashed-dotted line for variational
results). The number of atoms are plotted as functions of the
difference $\mu_-=\mu_1-\mu_2$ at fixed total chemical potential
$\mu_+=\mu_1+\mu_2$. The dotted straight lines in (a) and (e) correspond
to the $\mu_+$ of the particle number profiles given in (b) -(d) and
(f)-(g), respectively. } \label{N1N2sigmaPositive}
\end{figure}

\subsection{Attractive inter-component interactions ($\sigma>0$)}

The coupled differential equations (\ref{StationaryEqs}) were solved numerically by  
a relaxation method similar to that described in
Ref.~\cite{Petviashvili76}. Furthermore, we performed a variational
analysis of the fundamental vector solitons. The variational results
not only provide an appropriate initial condition for our numerical
relaxation procedure but also they show a good 
 agreement with our numerical results as is seen e.g.
from Figs.~\ref{N1N2sigmaPositive} (b)-(d) and (f)-(h).

If all interactions between particles are attractive, then both
solitonic components are basically bell-shaped and a simple trial
function of the form
\begin{equation}
\label{PsiTrial_Attractiv} \psi_j(r)=\sqrt{\frac{N_j}{\pi a_j^2}} e^{-\frac{1}{2}\frac{r^2}{a_j^2}}
\end{equation}
is expected to be a good approximation. Here $a_j$ is the effective
width of component $j$. The trial function is normalized to the
number of particles: $\langle\psi_j|\psi_j\rangle=N_j$. Substituting
the trial function (\ref{PsiTrial_Attractiv}) into Eq. (\ref{E}), we
obtain the energy
\begin{equation}
\label{EVariat}
E=E_1+E_2-\sigma E_{12}
\end{equation}
with
$$E_j=N_j\left(\frac{1}{a_j^2}+a_j^2-\frac{N_j}{4\pi
a_j^2}\right),~~E_{12}=\frac{N_1N_2}{\pi(a_1^2+a_2^2)}.$$ A soliton
solution corresponds to the stationary point of the total energy at
the fixed number of particles in  each component: $\left(\partial
E/\partial a_1\right)_{N_1,N_2}=0, \left(\partial E/\partial
a_2\right)_{N_1,N_2}=0.$

Obviously, the two-parameter family of vector soliton solutions is
symmetric with respect to the simultaneous interchange
$\mu_1\leftrightarrow\mu_2$ and $\psi_1\leftrightarrow\psi_2$, so that in
the plane of chemical potentials $(\mu_1,~\mu_2)$ the solutions are
symmetric with respect to the line $\mu_1=\mu_2$. To utilize this
symmetry we present the number of particles for each component as
functions of the difference $\mu_-=\mu_1-\mu_2$ at the fixed total
chemical potential $\mu_+=\mu_1+\mu_2$. Typical results are shown in
Fig. \ref{N1N2sigmaPositive} (b)-(d) and (f)-(h).

Note, that the region of existence for vector soliton pairs has
boundaries in the $(\mu_1,~\mu_2)$ plane, beyond which only
single-component (scalar) solitons exist. For different fixed values
of $\mu_+$ we determined the number of particles in each component
and then constructed the complete existence region, which is shown
in Figs. \ref{N1N2sigmaPositive} (a) and (e). In these figures the
solid black line separates the vector solitons from the scalar
soliton region where one solitonic component vanishes and the red
dashed line indicates the boundary where the other component
vanishes. As expected, these curves merge at the point
$\mu_1=\mu_2=2$ where both components vanish. With increasing
interaction strength $\sigma$ the region of existence for vector
solitons gradually shrinks, and at $\sigma=1$ the boundaries
degenerate to a straight line. Indeed, only scalar solitons with
equal chemical potential exist for $\sigma=1$. For $\sigma>1$, where
inter-component interactions dominate over intra-component
interactions, the region of existence for vector solitons grows
again. With respect to the case $\sigma<1$ the boundary lines of the
existence region  swap around [compare Fig. \ref{N1N2sigmaPositive}
(a) and Fig. \ref{N1N2sigmaPositive} (e)].

As is seen from the Fig.~\ref{N1N2sigmaPositive}, the number of
particles $N_j(\mu_-)$ reaches its maximum value $N_{j\,{\rm max}}$
at the boundary of the existence domain where the other component
$N_{3-j}$ vanishes. For decreasing total chemical potential $\mu_+$,
$N_{j\,{\rm max}}$ increases towards $N_{\rm cr}=11.68$ for
numerical solutions and $N_{\rm cr}=4\pi$ for the variational trial
functions~(\ref{PsiTrial_Attractiv}).

The remarkably different behavior of the functions $N_j(\mu_{-})$
for weak or strong inter-component interactions is evident from a
comparison of Figs.\ref{N1N2sigmaPositive}~(d) and (h). In the first
case ($0<\sigma<1$) the number of atoms saturates at $N_{j\,{\rm
max}}\le N_{\rm cr}$. In the second case ($\sigma>1$) the number of
atoms decays rapidly from $N_{j\,max}\le N_{cr}$ at one boundary to
zero at the other boundary of the existence domain. It is easy to
verify that for the case $\sigma>1$ the sum of the number of
particles in each component is always less than the critical value $N_{\rm cr}$.
For weak inter-component interactions ($0<\sigma<1$) we may find
$N_1+N_2>N_{\rm cr}$.

In summary, if inter-component as well as intra-component
interactions are attractive, then both solitonic components are
localized at the bottom of the potential trap, and the number of
atoms in each component is subcritical. These facts change
dramatically for repulsive inter-component interactions, which we
discuss next.

\subsection{Repulsive inter-component interactions ($\sigma<0$)}

To gain insight into the properties of the fundamental vector
solitons with inter-component repulsion we first present the results
of a variational analysis. They provide a rather complete overview
of the steady states which correspond to stationary points of the
total energy~(\ref{E}) at a fixed number of atoms in each component.

Minima of the total energy at a given number of atoms can be
attained for two different types of the solutions. The first type we
call ``bell-shaped" since the density distribution of both
components looks Gaussian-like. To stabilize such a ``miscible"
configuration for negative $\sigma$, the particle number in each
component must increase with respect to the previously discussed
case $\sigma>0$ in order to compensate for the additional
inter-component repulsion by increased intra-component attraction.
Solutions of the second type will be referred to as
``phase-separated" or ``immiscible". These solutions minimize the
``component-overlapping" part of the total energy [the last term in
Eq. (\ref{E})]  by substantially changing the shape of the density
distributions. One component is pushed outwards and forms a
ring-like shell, while the other component is noticeably compressed
—- it has a higher peak density and narrower width than its
non-interacting counterpart. The ``immiscible" states are typical
for BECs with both repulsive intra- and inter-component interactions
\cite{ground2,ground1,PuBigelow2,PuPRA09,Adhikari01,Zhou08}. The
notable feature of BECs with attractive intra-component but
repulsive inter-component interactions is the possible coexistence
of ``miscible" and ``immiscible" states for the same chemical
potentials.

\begin{figure}
\includegraphics[width=3.4in]{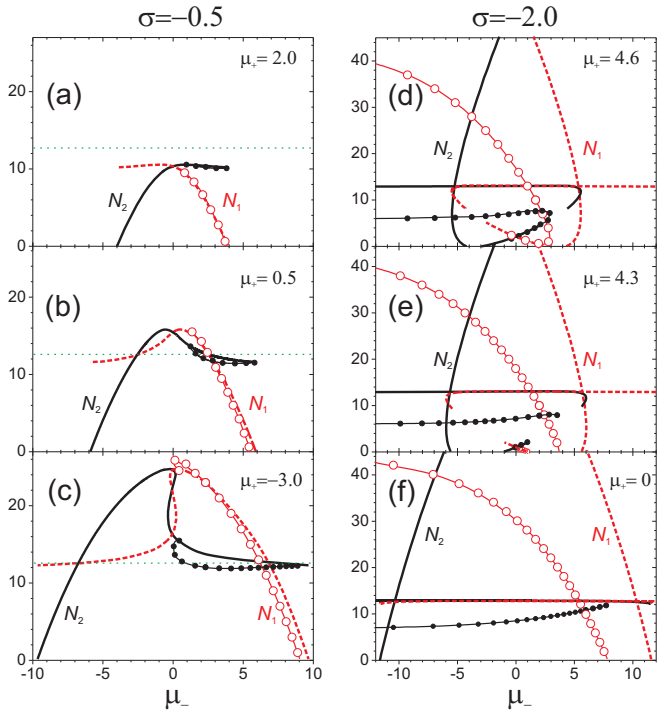}
\caption{(Color online) Variational results  for  $\sigma=-0.5$ [(a)-(c)] and
$\sigma=-2.0$ [(d)-(f)]. Shown are the number of atoms $N_1$
(red dashed curve for solutions with $\delta=0$,
red curves with open circles for solution with $\delta>0$) and $N_2$
(solid black curves for solutions with $\delta=0$ and black curves
with filled dots for solutions with $\delta > 0$) as the functions of $\mu_-=\mu_1-\mu_2$ at
different values of the total chemical potential $\mu_+=\mu_1+\mu_2$.
Green dotted line: $N=N_{\rm cr}=4\pi$.}
\label{VariatRepulsiveN1N2}
\end{figure}

\begin{figure}
\includegraphics[width=3.4in]{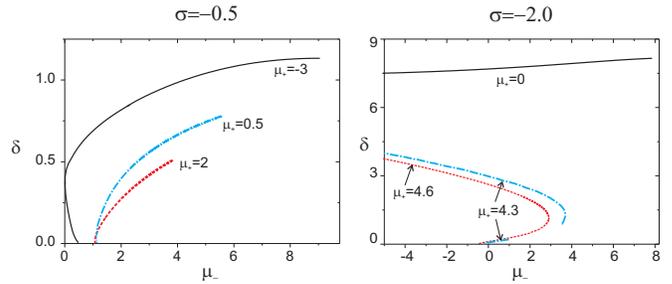}
\caption{(Color online) Variational parameter $\delta$ as a function of
$\mu_-=\mu_1-\mu_2$ for different values
of the total chemical potential $\mu_+=\mu_1+\mu_2$.} \label{deltaSigmaMinus}
\end{figure}

The simplest normalized trial function $\psi_j(r)$ supporting
spatially separated components is given by the Ansatz
\begin{equation}
\label{PsiTrial} \psi_j(r)=A_j\left(1+\delta_j \frac{r^2}{a^2_j}\right)
e^{-\frac{1}{2}\frac{r^2}{a_j^2}},
\end{equation}
where $a_j$ is the effective width of component $j$ of the
soliton-soliton pair, and $A_j=\sqrt{N_j/(\pi
a_j^2(1+2\delta_j(1+\delta_j)))}$. The parameter $\delta_j\ge 0$
introduces deviations from the Gaussian-like shape of the soliton.
If $\delta_j>1/2$, the density distribution has a local minimum at
the bottom of the external potential trap. Evidently only one of the
two BEC components is forced outwards by repulsive inter-component
interactions. Consequently, $\delta_2$ may be set to zero without
loss of generality and we need to determine only three variational
parameters: $a_1$, $a_2$ and $\delta=\delta_1$. Variational results
for the number of atoms as functions of the chemical potential
difference $\mu_-=\mu_1-\mu_2$ are shown in Fig.
\ref{VariatRepulsiveN1N2}.

Let us first discuss the properties of the ``bell-shaped" solutions
with $\delta=0$. These solutions are presented in Fig.
\ref{VariatRepulsiveN1N2} by the red dashed curves for $N_1$ and by
the solid black line for $N_2$. As it should, solutions with the
same chemical potential and same number of atoms only exist for
$\mu_+<4$ if $-1<\sigma<0$ and for $\mu_+>4$ if $\sigma<-1$. For
$-1<\sigma<0$, the existence domain of vector solitons is bounded in
the $(\mu_1,~\mu_2)$ plane like in the case $\sigma>0$.

However, if inter-component repulsion dominates over intra-component
attraction ($\sigma<-1$) ``bell-shaped" solutions exist for any
chemical potential. It is remarkable that even at $\mu_-=0$ the
solitonic components differ essentially from each other for
$\mu_+<4$ [see Fig. \ref{VariatRepulsiveN1N2} (f)], but for
$\mu_+>4$ there appears an additional branch of solutions that share
the same profile at $\mu_-=0$ [see Fig. \ref{VariatRepulsiveN1N2}
(d),(e)]. The gap between these two branches gradually disappears
when $\mu_+$ increases [compare Figs. \ref{VariatRepulsiveN1N2} (e)
and (d)]. For relatively weak inter-component interactions
($-1<\sigma<0$) and decreasing $\mu_+$ the absolute value of the
derivatives $|\partial N_j/\partial \mu_j|$ grows rapidly in the
vicinity of the intersection point. Furthermore, it is interesting,
that in the profile for $N_j(\mu_-)$ a fold appears for
$N_j(0)>16\pi/(3+\sigma)$ provided that $-1<\sigma<-1/3$. In this
case two additional crossing points are observed at $\mu_1=\mu_2$
[see Fig. \ref{VariatRepulsiveN1N2} (c)].

\begin{figure}
\includegraphics[width=3.4in]{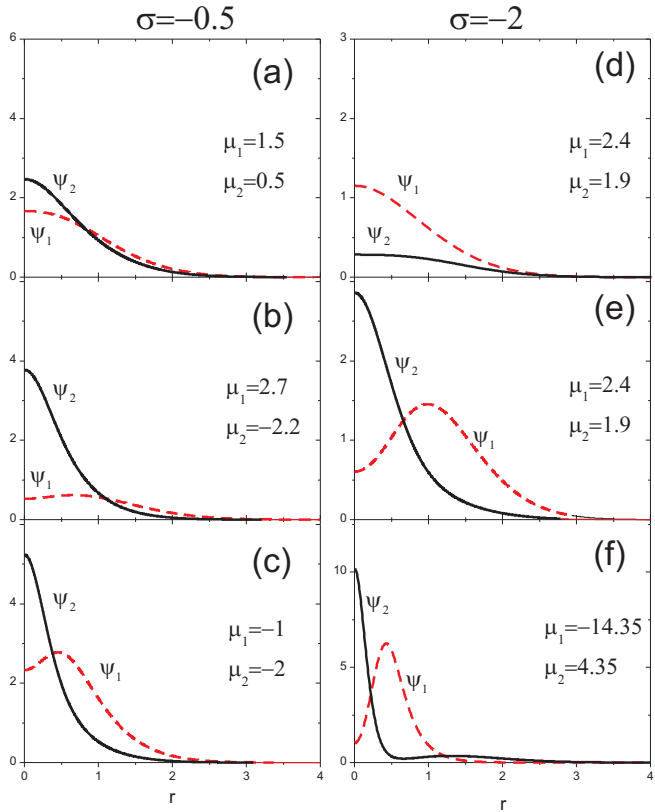}
\caption{(Color online) Examples of radial profiles $\psi_1$ (red dashed curves)
and $\psi_2$ (black solid curves) for $\sigma=-0.5$ and  $\sigma=-2.0$
found numerically.} \label{RadialProfiles}
\end{figure}

\begin{figure}
\includegraphics[width=3.4in]{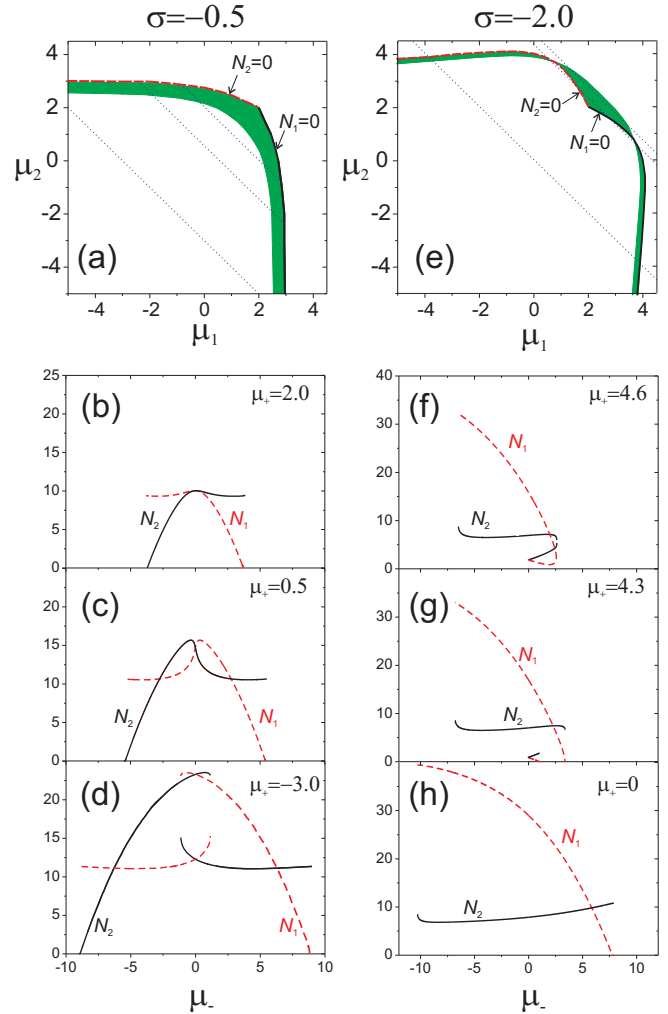}
\caption{(Color online) Top: Stability region in the $(\mu_1,~\mu_2)$ plane (green)
as obtained by numerical simulations: (a) $\sigma=-0.5$, (b) $\sigma=-2.0$.
Bottom: Number of atoms $N_1$ (red dashed curves) and $N_2$ (black solid curves) as
a function of the chemical potential difference $\mu_-=\mu_1-\mu_2$ at fixed total
chemical potential $\mu_+=\mu_1+\mu_2$ obtained numerically: (c), (d), (e) $\sigma=-0.5$,
(f), (g), (h) $\sigma=-2.0$. The dotted straight lines in (a) and (e) correspond
to the $\mu_+$ of the particle number profiles given in (b) -(d) and
(f)-(g), respectively.}
\label{SteadySigmaNegative}
\end{figure}

The properties of the second type of solutions (``phase-separated"
states with $\delta>0$), predicted by the variational method, are
illustrated in Fig. \ref{VariatRepulsiveN1N2} for   $N_1(\mu_-)$ by
red lines with open circles, and for $N_2(\mu_-)$ by black curves
with filled dots. Recall that we show only solutions with
$\delta_1=\delta$ and $\delta_2=0$ but the rest of  solutions can be
easily constructed using the corresponding symmetry. The parameter
$\delta$ describing the deviation of the radial profile from the
Gaussian-like shape is given in Fig. \ref{deltaSigmaMinus}. Here,
$\delta(\mu_-)$ is shown for the same values of $\sigma$ and $\mu_+$
as in Fig. \ref{VariatRepulsiveN1N2}. As is seen from Fig.
\ref{deltaSigmaMinus} the variational parameter $\delta$ reaches
zero at some value of $\mu_-$, where the two types of solutions
$\delta>0$ and $\delta=0$ merge. Indeed this is also observed in
Fig. \ref{VariatRepulsiveN1N2} for solutions with
$\mu_1\approx\mu_2$.

The above described features of phase-separated steady-states
obtained using the variational approach are supported by numerical
simulations of Eqs. (\ref{StationaryEqs}). Note that only half of
the phase-separated solutions are shown in Figs.
\ref{SteadySigmaNegative} (f)-(h) to avoid confusion. The traces in
these figures can easily be completed using the replacements $N_1\to
N_2$ and $\mu_-\to-\mu_-$. Unfortunately, our relaxation technique
is not able to reproduce both types of solutions predicted by
variational method for repulsive inter-component interactions since
the ``bell-shaped" solutions (with $\delta_1=\delta_2=0$) usually
have higher energy than the phase-separated solutions at the same
chemical potential. That is why our numerical procedure rather
converges to the ``phase-separated" solutions or to scalar solitons.
Therefore, our analysis of the steady-states with
$\delta_1=\delta_2=0$ is 
 limited to the
variational method.

Typical radial profiles found numerically are shown in Fig.
\ref{RadialProfiles}. Generally, the numerical results are found to
be in good agreement with the variational predictions for the
``phase-separated" solutions. One expects, however, differences
between numerical and variational results for solutions with a very
dense ring-shaped component. Indeed, a ring component with large
amplitude could push away the atoms at the periphery of the core
component. Consequently, an additional outer ring should be observed
in the density distribution of the core component. Moreover, if the
number of atoms increases even more, then more and more rings should
appear in both solitonic components. Actually, an additional outer ring is seen for $\psi_2(r)$ in Fig. \ref{RadialProfiles} (f), and such
a modification of the core component explains the non-monotonic
behavior of the diagrams $N_2(\mu_-)$ at negative $\mu_-$ in Figs.
\ref{SteadySigmaNegative} (f)-(h).

\section{Stability analysis}

A few general conclusions concerning stability follow from our
variational analysis. In fact, it is easy to find a
\textit{sufficient} condition for stability against radially-symmetric collapse: if both solitonic
components satisfy the condition $N_j<N_{\rm cr}$, then the
stationary solutions are expected to be stable since they correspond
to the minimum of the total energy. As was shown in the previous
section, for attractive inter-component interactions ($\sigma>0$)
the particle numbers in both solitonic components are always below
the critical value and, therefore, the vector solitons should be
stable in this case.

For $\sigma<0$, the variational analysis predicts the existence of
two types of stationary solutions: ``phase-separated" states and
colocated ``bell-shaped" states with different stability properties.
The solutions with $\delta_1=\delta_2=0$ should be stable provided
$N_1<N_{\rm cr}$ and $N_2<N_{\rm cr}$. At the same time, the
phase-separated states are predicted to be stable against collapse
even if the ring-component $j$ (with $\delta_{j}>0$) is
over-critical  provided that the bell-shaped component (with
$\delta_{3-j}=0$) consists of an undercritical number of particles.

It is clear that a radially-symmetric variational analysis does not
provide any information about the stability with respect to
azimuthal perturbations. This question may be addressed by numerical
simulations as discussed below. In contrast to the results of the
variational analysis  our numerical simulations predict the
phase-separated states to be unstable against collapse even if only
one component has a number of particles above the critical value
$N_{\rm cr}$. Note that numerical time evolution of the variational
Ansatz for the steady-states with $\delta_1=\delta_2=0$ confirmed
their stability only in the region where these solutions practically
merge with the second solitonic branch [see Figs.~
\ref{VariatRepulsiveN1N2}, \ref{deltaSigmaMinus}, and
\ref{SteadySigmaNegative}].

The stability of the stationary solutions with respect to azimuthal
symmetry-breaking perturbations was investigated by a linear
stability analysis.  We write the order parameter of the
two-component BEC in the form
$\tilde\Psi_j(r,t)=(\psi_j(r)+\varepsilon_j(r,t))e^{-\mu_jt}$ with a small
perturbation  $\varepsilon_j(r,t)=u_j(r)e^{i\omega t+i
L\varphi}+v^*_j(r)e^{-i\omega^* t-i L\varphi}$ and insert it into
Eq. (\ref{StationaryEqs}). We linearized the resulting equations
with respect to $\varepsilon$ and solved the resulting eigenvalue
problem for $\omega$. Solutions exhibiting an imaginary part would
indicate instability.

\begin{figure}
\includegraphics[width=3.4in]{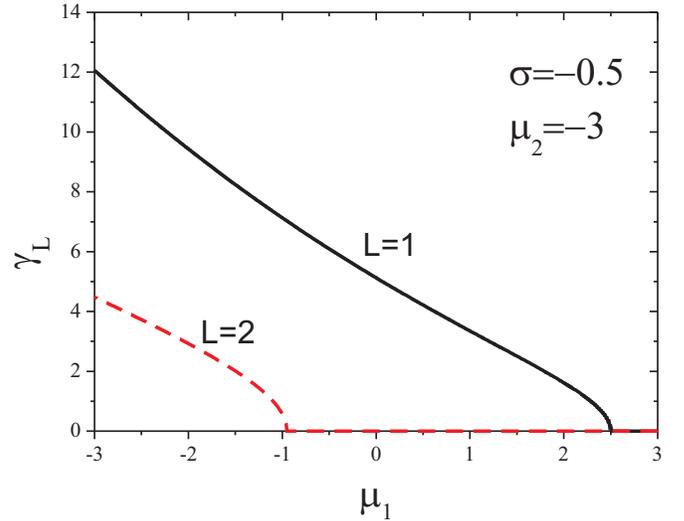}
\caption{(Color online) Maximum growth rates as a function of $\mu_1$ at fixed $\mu_2$.
Note that all growth rates vanish above some
threshold value of $\mu_1$.} \label{GrowthRates}
\end{figure}

We find that  $\gamma_L=\textrm{Im}(\omega)$ are identically zero for all
$L$ if the inter-component interaction is attractive. If it is
repulsive the azimuthal modes $L=1$ and $L=2$ may be unstable. A
typical example for a maximum growth rate $\gamma_L$ as a function
of the chemical potential $\mu_1$ at fixed $\mu_2$ is shown in
Fig.~\ref{GrowthRates}. Note that above some threshold the growth
rates vanish. It is important to realize that the mode $L=1$ has a
much wider instability region than $L=2$, and that the growth
rates of the mode $L=2$ are smaller than those of the mode $L=1$.
Therefore, the  $L=1$ modes are most dangerously influencing the
stability of repulsively interacting solitons.

\begin{figure}[t]
\includegraphics[width=3.4in,height=8.0in,keepaspectratio]{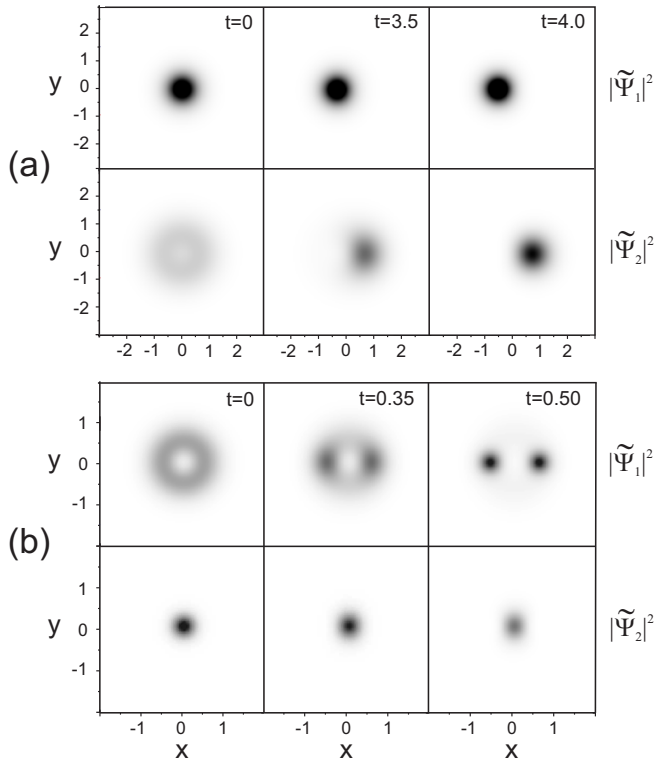}
\caption{Time evolution of the density distributions $|\tilde\Psi_1|^2$ (upper rows)
and $|\tilde\Psi_2|^2$ (lower rows)  of perturbed vector solitons for (a)
$\sigma=-0.5$, $\mu_1=-1.5$, $\mu_2=2$ and (b) $\sigma=-0.8$, $\mu_1=-5$, $\mu_2=-5$.}
\label{Dynamics}
\end{figure}

The predictions of the variational analysis and the results of the
linear stability analysis have been tested by numerical experiments
of the time evolution of the perturbed vector solitons. For the
numerical solution of the time-dependent GPEs (\ref{GP2D}) a
standard split-step Fourier transform method (see, e.g., Ref.
\cite{Agrawal}) has been used. While vector solitons with attractive
inter-component interactions are confirmed to be stable over the
full existence domain, repulsive inter-component interactions lead
to various instabilities. Typical examples for unstable dynamics are
given in Fig. \ref{Dynamics}.

We found that, if at least one solitonic component contains an
overcritical number of atoms, then the stronger component finally
collapses. However, we do not see a collapse, if both solitonic
components are under-critical. Obviously, the potential trap
prevents the unlimited expansion of the BEC. Furthermore, the final collapse is not
possible, if the number of atoms in both components is subcritical, so as the result of symmetry-breaking instability we see a
contraction and relative motion of the two solitonic components. Oscillations between
such a quasi-collapse and an asymmetric state is shown in
Fig.~\ref{Dynamics}~(a) for times $t=3.5$ and $t=4.0$. Here, both
solitonic components are under-critical, and the growth rate for the
$L=1$ mode is non-vanishing.

The growth of the $L=2$ mode leads to the decay of the shell
component of the soliton into two filaments. At the beginning, two humps and two
holes appear in the ring, and the inner component rapidly leaks
through these hollows into a ring-shaped ``well". Repulsive
inter-component interactions between the ring-shaped component and
the leaking inner component leads to an increase of the hollows and
finally the ring breaks up into two separate parts as seen in
Fig.~\ref{Dynamics}~(b). Note that the dynamics of the unstable
trapped two-component BEC is similar to the evolution of vector
solitons reported in Ref.~\cite{PRE2010} within a similar model but
without external trapping potential. The most outstanding feature of
trapped matter-wave vector solitons is the fact that they can be
completely stabilized even for strong repulsive inter-component
interactions. We show the stability region in green color in the
$(\mu_1,~\mu_2)$ plane for weak ($\sigma=-0.5$) and strong
($\sigma=-2.0$) inter-component repulsive interactions in
Figs.~\ref{SteadySigmaNegative}~(a) and (e).

\section{Summary and conclusions}
Fundamental 2D soliton-soliton pairs are investigated in
two-component BECs with attractive intra-component interactions.
General properties of vector solitons and their stability are
studied variationally and numerically for both attractive and
repulsive inter-component interactions. We found different types of
soliton-soliton pairs including phase-separated pairs where one
component is pushed outwards and forms a ring-like shell and the
other component is compressed due to repulsive inter-component
interactions. It turns out that for some values of the chemical
potentials phase-separated steady-states coexist with collocated
states characterized by bell-shaped density distributions in both
components.

We performed a linear stability analysis of small azimuthal
perturbations and checked these results by an extensive series
of numerical simulations. For attractive inter-component
interactions matter-wave bright vector solitons are demonstrated to
be stable throughout the existence domain. For BEC components which
repel each other various unstable evolution scenarios including
collapse and azimuthal symmetry-breaking instabilities are observed.
The instabilities, as a rule, lead either to separation of the
condensed phases and then a collapse of the stronger over-critical
($N>N_{\textrm cr}$) component or a periodic relative motion of the
under-critical ($N<N_{\textrm cr}$) solitonic components backwards and
forwards near the bottom of trapping potential. Nevertheless, there
are conditions where complete stabilization of vector solitons is
observed even in the case of repulsive inter-component interactions.

\section{Acknowledgment}
The work was supported by the grants WE1085/5-1 (DFG, Germany),
A/11/05264 (DAAD, Germany), and F39.2 (SFFR, Ukraine). We thank V.M.
Lashkin for useful discussions.

\bibliography{refs}

\end{document}